\documentstyle[pre,aps,epsf]{revtex}
\newcommand{\sgn}{\text{sgn}}
\newcommand{\J}{\mbox{\bf J}}
\newcommand{\ksi}{\bbox{\xi}}

\begin{document}
\draft
%++++++++++++++++++++++++++++++++++++++++++++++++++++++++++++++++++++++++++++
\title{Correlations between hidden units in multilayer neural networks and
replica symmetry breaking}
\author{D. Malzahn and A. Engel}
\address{Institut f\"ur Theoretische  Physik,
         Otto-von-Guericke-Universit\"at\\
         Postfach 4120, 
         D-39016 Magdeburg, Federal Republic of Germany}

\date{19 February 1999}

%++++++++++++++++++++++++++++++++++++++++++++++++++++++++++++++++++++++++++++
\maketitle
\begin{abstract}
We consider feed-forward neural networks with one hidden layer, 
tree architecture and a fixed hidden-to-output Boolean function. 
Focusing on the saturation limit of the storage problem the
influence of replica symmetry breaking on the 
distribution of local fields at the hidden units is investigated. These 
field distributions determine the probability for 
finding a specific activation pattern of the hidden units as well as the 
corresponding correlation coefficients and therefore quantify the division 
of labor among the hidden units.
We find that although modifying the storage capacity and the distribution of 
local fields markedly replica symmetry breaking has only a minor effect on 
the correlation coefficients. Detailed numerical results are provided for 
the {\small PARITY}, {\small COMMITTEE} and {\small AND} machines with K=3 
hidden units and 
nonoverlapping receptive fields. 
\end{abstract}


%++++++++++++++++++++++++++++++++++++++++++++++++++++++++++++++++++++++++++++
\section{Introduction}
%++++++++++++++++++++++++++++++++++++++++++++++++++++++++++++++++++++++++++++

Multilayer neural networks (MLN) are more powerful devices for information 
processing than the single-layer perceptron because of the 
possibility of {\it different} activation patterns, so-called internal 
representations (IR), at the hidden units for the {\it same} input-output 
mapping. 
It is well known that the correlations between the activities at the hidden 
units are crucial for the understanding of the storage and generalization 
properties of a MLN \cite{MePa,GriGro,Priel,Scho,MoZe,En96}. A particular 
simple situation to study these correlations is the implementation of random 
input-output mappings by the network, the so-called storage problem, near the
storage capacity. Using the replica trick and assuming replica symmetry the 
correlation coefficients building up in this case were calculated in 
\cite{En96} and shown to be characteristic for the prewired Boolean function 
between hidden layer and output. Conversely, {\it prescribing} these 
correlations the storage properties of the networks change \cite{DM}.

The assumption of replica symmetry (RS) in this calculation is somewhat 
doubtful. In fact it is well known that the storage capacity of MLN is 
strongly
modified by replica symmetry breaking (RSB) \cite{par,Barkai,En92}, which is 
due to the very possibility of different internal representations. Moreover, 
even the distribution of the output field of a simple perceptron is  
influenced by RSB effects \cite{PMaj,GyRe}. 

In the present paper we elucidate the impact of RSB on the correlation 
coefficients between the activity of different hidden units in MLN 
with one hidden layer and nonoverlapping receptive fields. 
The central quantity of interest is the joint 
probability distribution for the local fields at the hidden 
units. In the general part of this paper we show how this distribution can 
be 
calculated both in RS and in one-step RSB. For a detailed analysis we than 
specialize to MLN with $K=3$ hidden units   
and discuss, in particular, the {\small PARITY}, {\small COMMITTEE} and 
{\small AND} machines. 
Together with the corrections from one-step RSB the RS 
results give insight in the division of labor between different subperceptrons 
in MLN and the role of RSB. Calculating finally the correlation coefficients 
we find that although modifying the local field distribution markedly RSB 
gives rise to minor corrections to the correlation coefficients only.

%++++++++++++++++++++++++++++++++++++++++++++++++++++++++++++++++++++++++++++
\section{General results}
%++++++++++++++++++++++++++++++++++++++++++++++++++++++++++++++++++++++++++++
We consider feed-forward neural networks with $N$ inputs $\ksi_k^{\nu}$, 
one hidden layer of $K$ units $\tau_1,\tau_2,\dots,\tau_K$ and a single 
output $\sigma$. The hidden units have nonoverlapping 
receptive fields of dimension $N/K$ (tree structure). They are determined by 
the inputs via spherical coupling vectors $\J_k\in {\bf I\!R}^{N/K}, 
\J_k^2=N/K$ according to $\tau_k=\sgn(h_k)$ with
$h_k=\J_k\ksi_k\sqrt{K/N}$ denoting the local fields.
We call an activation pattern 
$(\tau^{\nu}_1,\tau^{\nu}_2,\dots,\tau^{\nu}_K)$ of 
the hidden units an internal representation (IR).
The output $\sigma$ of the MLN is a fixed Boolean function 
$\sigma=F(\tau_1,\dots,\tau_K)$ of the IR. Examples of special interest 
include the 
{\small PARITY} machine, $F(\{\tau^{\nu}_k\})=\prod_{k=1}^K \tau^{\nu}_k$, the 
{\small COMMITTEE} machine, $F(\{\tau^{\nu}_k\})=\sgn (\sum_{k=1}^K \tau^{\nu}_k)$, 
and the {\small AND} machine, $F= +1\:{\rm if\: all}\:\tau_k=+1;\:{\rm else}\: 
F= -1$.

All IR consistent with a desired 
output are called legal internal representations (LIR).
The number of and similarity between LIR to a given output specifies the 
division of labor taking place between the different perceptrons forming
the MLN. It is quantitatively characterized by the correlation coefficients
\begin{eqnarray}
c_n&=& 
\langle \langle \sigma\tau_{i_1}\tau_{i_2}\cdots\tau_{i_n} \rangle \rangle 
\label{correl},
\end{eqnarray}
$n=1,\dots,K$, where $\langle\langle \cdots \rangle\rangle$ denotes the 
average over the inputs and the output and $i_1,\dots,i_n$ is a subset of 
$n$ natural numbers between $1$ and $K$. For permutation symmetric Boolean 
functions, the $c_n$ only depend on $n$ and not on the particular choice of 
this subset. 

We focus on the so-called storage problem in which the inputs $\ksi_k^{\nu}$
and the outputs $\sigma^{\nu}$ are generated independently at random
according to the probability distributions
\begin{eqnarray}
p(\sigma^{\nu})=\frac{\delta(\sigma^{\nu}-1)+\delta(\sigma^{\nu}+1)}{2}
\end{eqnarray}
and
\begin{eqnarray}
p(\xi_{k,i}^{\nu})=
\frac{1}{\sqrt{2\pi}}\exp\left(-\frac{1}{2}(\xi_{k,i}^{\nu})^2\right),
\end{eqnarray}
where $k=1,\dots,K$, $i=1,\dots,N/K$ and $\nu=1,\dots,\alpha N$.

The basic quantity which gives us access to the probability of the LIR
and to the correlation coefficients is the
distribution $p(h_j)$ of the local fields  
$h_j$ at the $j$th hidden unit. It is given by
\begin{eqnarray}
p(h_j)=\bigg\langle\!\!\bigg\langle \frac{1}{\cal Z}
\int \prod\limits^{K}_{k=1}d\mu(\J_k)d\mu(\lambda^{\nu}_k)
\prod\limits^{\alpha N}_{\nu=1}
\Theta\left(\sigma^{\nu}F(\sgn(\lambda^{\nu}_1),\dots,\sgn(\lambda^{\nu}_K))
\right)
\delta\left(h_j-\lambda^1_j\right)
\bigg\rangle\!\!\bigg\rangle_{\{\ksi_k^{\nu}\},\sigma^{\nu}}.
\label{p_h}
\end{eqnarray}
$\langle\langle\cdots\rangle\rangle$ denotes the average over all stored 
input-output patterns.
${\cal Z}$ denotes the partition function 
\begin{eqnarray}
{\cal Z}&=&\int \prod\limits^{K}_{k=1}d\mu(\J_k)d\mu(\lambda^{\nu}_k)
\prod\limits^{\alpha N}_{\nu=1}
\Theta\left(\sigma^{\nu}F(\sgn(\lambda^{\nu}_1),\dots,\sgn(\lambda^{\nu}_K))
\right),
\end{eqnarray}
$d\mu(\J_k)$ the measure on the Gardner sphere \cite{Gard}
\begin{eqnarray}
d\mu(\J_k)&=&
\delta\left(\J_k^2 - \frac{N}{K}\right)\frac{d\J_k}{\sqrt{2\pi e}^{N/K}},
\label{measure_J}
\end{eqnarray}
and $d\mu(\lambda^{\nu}_k)$ the integration measure
\begin{eqnarray}
d\mu(\lambda^{\nu}_k)&=&
\delta\left(\lambda^{\nu}_k-\J_k\ksi^{\nu}_k\sqrt{\frac{K}{N}}\right)
d\lambda^{\nu}_k.
\label{measure_fields}
\end{eqnarray}

We use the replica trick $1/{\cal Z}=\lim_{n\rightarrow 0}{\cal Z}^{n-1}$ 
in Eq. (\ref{p_h}) to perform the average over the inputs $\{\ksi^{\nu}_k\}$
and introduce the overlaps $q_k^{ab}=\J_k^a\J_k^b/(N/K)$ 
between different replicas $a,\:b$ of a coupling vector $\J_k$
of hidden unit $k$.
We will consider only permutation symmetric Booleans $F$. 
Hence all hidden units have the same statistical properties implying 
$p(h_k)=p(h)$ and $q_k^{ab}=q^{ab}$ with $k=1\dots K$.
Equation (\ref{p_h}) takes on the form
\begin{eqnarray}
p(h_j)=\lim\limits_{n\rightarrow 0}
\int\prod\limits_{a<b} dq^{ab}\:
\langle\!\langle p(h_j|\sigma)\rangle\!\rangle_{\sigma}
\:\:\exp\left(\frac{N}{2}
\ln {\rm det}(Q)
+(\alpha N-1)\langle\!\langle\ln G_1(Q|\sigma)\rangle\!\rangle_{\sigma}
\right)
\label{ph_1}
\end{eqnarray}
in terms of the $(n\times n)$-dimensional order parameter matrix $Q$ where  
$Q^{aa}=1$ and $Q^{ab}=q^{ab}$. Here 
\begin{eqnarray}
p(h_j|\sigma)&=&{\displaystyle \int}\prod_{k,a} 
\frac{d\lambda_k^{a}dx_k^{a}}{2\pi}
\exp\left(\sum_{k,a}\left[i\;x_k^{a}\lambda_k^{a}-\frac{1}{2}
(x_k^{a})^2\right]
-\sum_{k,a<b}x_k^{a}x_k^{b}q^{ab}\right)
\nonumber\\
&\times&\prod\limits_{a}\Theta(\sigma
F(\{\sgn(\lambda_k^{a})\}))\delta(h_j-\lambda_1^{1})
\label{ph_central},
\end{eqnarray}
and the expression for $G_1(Q|\sigma)$ is specified in 
the Appendix, Eq. (\ref{G1}), together with some more details of the calculation.

In the limit $N\rightarrow \infty$ the integral (\ref{ph_1}) is dominated by 
the saddle point values of the order parameters $q^{ab}$ which 
extremize the partition function
\begin{eqnarray}
{\cal Z}=\exp\left(N \:{\rm extr}_{q^{ab}}\left\{\lim\limits_{n\rightarrow 0}
\frac{\frac{1}{2}\ln {\rm det}(Q)
+\alpha \langle\!\langle\ln G_1(Q|\sigma)\rangle\!\rangle_{\sigma}}{n}
\right\}\right)\quad.
\label{sad_cond}
\end{eqnarray}

In the following, we simplify Eqs. (\ref{ph_central}) and (\ref{sad_cond}) 
using the assumption 
that the order parameter matrix $Q$ is either replica symmetric or 
describes one-step replica symmetry breaking.
We will always consider the saturation limit $\alpha\to\alpha_c$ since the 
expressions then simplify and the correlations 
become most characteristic in this limit.  
The RS case is specified by \cite{Parisi}
\begin{eqnarray}
q^{ab}=\left\{
\begin{tabular}{cl}
$1$   & {\rm if} $a=b$\\
$q$   & {\rm else}.
\end{tabular}
\right.
\label{rs}
\end{eqnarray}
The saturation limit $\alpha\rightarrow \alpha_c$ is characterized by 
the existence of a unique solution $\J_k$, e.g., $q\rightarrow 1$.
We then get 
\begin{eqnarray}
p(h|\sigma)=
\int\prod\limits_{k=1}^K Dy_k\: \lim\limits_{q\rightarrow 1}\left(
\frac{\exp\left(-\frac{1}{2}(h+y_1\sqrt{q})^2/(1-q)\right)
}{\sqrt{2\pi(1-q)}}
\:
\frac{\Phi_{\rm LIR}\left(\sigma\:|\:\delta_{\eta_1,\sgn(h)}\right)
}{\Phi_{\rm LIR}(\sigma)}\right)
\label{ph_rs}
\end{eqnarray}
for the conditional probability to find a specific value $h$ of the 
postsynaptical potential under the constraint of a given output $\sigma$.
The terms abbreviated by
\begin{eqnarray}
\Phi_{\rm LIR}\left(\sigma\:|\:\delta_{\eta_1,\sgn(h)}\right)&:=&\!\!\!\!\!\!
\sum\limits_{{\rm all\: sets}\: (\eta_1,\dots,\eta_K)}\!\!
\delta_{\eta_1,\sgn(h)}\:
\delta_{\sigma,F(\eta_1,\dots,\eta_K)}
\prod\limits_{k=2}^K H\!\!\left(\eta_k y_k\sqrt{\frac{q}{1-q}}\right),
\label{LIR1_rs}\\
\Phi_{\rm LIR}(\sigma)&:=&\!\!\!\!\!\!
\sum\limits_{{\rm all\: sets}\: (\eta_1,\dots,\eta_K)}\!\!
\delta_{\sigma,F(\eta_1,\dots,\eta_K)}
\prod\limits_{k=1}^{K}H\!\!\left(\eta_ky_k\sqrt{\frac{q}{1-q}}\right)
\label{LIR2_rs}
\end{eqnarray}
ensure that only LIR for the respective value of $\sigma$
contribute to the sum in Eq. (\ref{ph_rs}). As usual we have used the 
error function $H(x)=\int_x^{\infty}Dt$ with $Dt=\exp(-t^2/2)dt/\sqrt{2\pi}$.
 
Let us now turn to main features of the solution within the ansatz of one-step 
RSB. Then the following form for the order parameter 
matrix is assumed \cite{Parisi}:
\begin{eqnarray}
q^{ab}=\left\{
\begin{tabular}{cl}
$1$   & {\rm if} $a=b$\\
$q_1$ & {\rm if} $|a-b|<m$\\
$q_0$ & {\rm else}.
\end{tabular}
\right.
\label{rsb}
\end{eqnarray}
Accordingly there are two overlap scales characterizing the similarity 
between
coupling vectors belonging to the same and different regions of the solution 
space, respectively.

Using this ansatz we find after standard manipulations \cite{Parisi}
for the probability distribution of the local field for 
a specific output $\sigma$ 
\begin{eqnarray} 
p(h|\sigma)=
\int\prod\limits_{k=1}^K Dy_k 
\frac{\displaystyle
{\displaystyle \int}\prod\limits_{k=1}^K Dz_k\:
\frac{\displaystyle 1}{\sqrt{\displaystyle 2\pi(1-q_1)}}
\exp\left(-\frac{\displaystyle (h+y_1\sqrt{q_0}+z_1\sqrt{q_1-q_0})}
{\displaystyle 2(1-q_1)}^{\!\!2}
\right)
\:\frac{\displaystyle
\Phi_{\rm LIR}\left(\sigma|\delta_{\eta_1,\sgn(h)}\right)}
{\displaystyle\left(\Phi_{\rm LIR}(\sigma)\right)^{1-m}}
}{\displaystyle
{\displaystyle \int}\prod\limits_{k=1}^K Dz_k \left(\displaystyle
\Phi_{\rm LIR}(\sigma)
\right)^m
},
\label{ph_rsb}
\end{eqnarray}
where now 
\begin{eqnarray}
\Phi_{\rm LIR}\left(\sigma|\delta_{\eta_1,\sgn(h)}\right):=
\!\!\!\!\!\!\sum\limits_{{\rm all\: sets\:}(\eta_1,\dots,\eta_K)}\!\!
\delta_{\eta_1,\sgn(h)}\:
\delta_{\sigma,F(\eta_1,\dots,\eta_K)}
\prod\limits_{k=2}^K 
H\!\!\left(\eta_k\frac{y_k\sqrt{q_0}+z_k\sqrt{q_1-q_0}}{\sqrt{1-q_1}}\right),
\label{sum_LIR_2}
\end{eqnarray}
\begin{eqnarray}
\Phi_{\rm LIR}(\sigma):=\!\!\!\!\!\!
\sum\limits_{{\rm all\: sets\:}(\eta_1,\dots,\eta_K)}\!\!
\delta_{\sigma,F(\eta_1,\dots,\eta_K)}
\prod\limits_{k=1}^{K}
H\!\!\left(\eta_k\frac{y_k\sqrt{q_0}+z_k\sqrt{q_1-q_0}}{\sqrt{1-q_1}}\right).
\label{sum_LIR_1}
\end{eqnarray}
These expressions simplify in the saturation limit
$\alpha\rightarrow\alpha_c$ in which one finds $q_1\rightarrow 1$ and 
$m=w(1-q_1)\rightarrow 0$.
The remaining order parameters $w,\:q_0$ are given by the saddle point 
equations  
corresponding to the following expression for the storage capacity $\alpha_c$:
\begin{eqnarray}  
\alpha_c=\min\limits_{q_0,w}
\left[\frac{\ln[1+w(1-q_0)] + q_0 w/[1+w(1-q_0)]}
{-2\lim\limits_{q_1\rightarrow 1}
\left\langle\!\!\left\langle{\displaystyle \int}\prod\limits_{k}Dy_k\ln\left\{
{\displaystyle \int}\prod\limits_{k}Dz_k\left(\Phi_{\rm LIR}(\sigma)\right)^m\right\}
\right\rangle\!\!\right\rangle_{\sigma}
}\right]\quad .
\label{saddle2}
\end{eqnarray}
As in the RS case the analytical and numerical analysis of these 
expressions for concrete situations needs some care (see next section).

To finally obtain $p(h)$ we must average Eqs. (\ref{ph_rs}) and 
(\ref{ph_rsb}) over the 
two possible outputs $\sigma=\pm 1$,
\begin{eqnarray}
p(h)=\langle\langle\: p(h|\sigma) \:\rangle\rangle_{\sigma}\quad .
\label{ph_rs_av}
\end{eqnarray}
From this probability distribution we find 
the distributions $p(\tau_1,\dots,\tau_K)$ of the LIR according to
\begin{eqnarray}\label{h1}
p(\tau_1,\dots,\tau_K)\!\!&=&\!\!\int\limits_{-\infty}^{\infty}\!
\prod\limits_{k=1}^{K}dh_k\: 
\Theta(\tau_k\: h_k)\:p(h_k).
\end{eqnarray}
The correlation coefficients $c_n$, $n=1,\dots,K$, are then given by
\begin{eqnarray}
c_n=\sum\limits_{{\rm all\: sets\:}(\eta_1,\dots,\eta_K)}
\sigma\:\eta_{1}\eta_{2}\cdots\eta_{n}\:
\delta_{\sigma,F(\eta_1,\eta_2,\dots,\eta_K)}\:
p(\eta_1,\eta_2,\dots,\eta_K).
\label{corr1}
\end{eqnarray}
The Kronecker $\delta$ in Eq. (\ref{corr1}) restricts the sum to all LIR 
of the output $\sigma$. 
Equation (\ref{corr1}) is valid as long as the pattern load
of the MLN does not exceed its saturation threshold $\alpha_c$.

\section{Specific examples with $K=3$ hidden units}
In this section we apply the general formalism developed above to the 
analysis 
of simple versions of three popular examples of MLN, namely, the {\small PARITY}, 
{\small COMMITTEE} and {\small AND} machines, each with $K=3$ hidden units. We start with the 
RS results.

\subsection{Replica symmetry}
In {\small COMMITTEE} and {\small PARITY} machines there is for every LIR of output 
$\sigma=+1$ 
an IR with all signs reversed that realizes output $\sigma=-1$. 
Therefore $p(h)=p(h|+1)=p(h|-1)$ and the final average over $\sigma$ in 
Eq. (\ref{ph_rs_av}) is 
trivial. Analyzing Eqs. (\ref{LIR1_rs}) and (\ref{LIR2_rs}) in the limit
$q\rightarrow 1$ one realizes that they depend on both the sign and values 
of 
all integration variables $y_k$. Expression (\ref{LIR1_rs}) as well as 
Eq. (\ref{LIR2_rs}) are either equal to one or exponentially small in some or 
all 
integration variables. The quotient of both figuring in Eq. (\ref{ph_rs})
can hence become one, zero, or singular with respect to $y_1$.
Whenever it is one the integral in 
Eq. (\ref{ph_rs}) gives rise to $\delta(h+y_1)$ for $q\to1$. 
Whenever the quotient is singular a contribution $\delta(h)$ results.

Keeping track of the different contributions arising in this way we find
for the $K=3$ {\small COMMITTEE} machine
\begin{eqnarray}
p(h)= \Theta(-h)\frac{e^{-h^2/2}}{\sqrt{2\pi}}H^2(h) + 
\frac{5}{24}\delta_{+}(h) + \Theta(h)\frac{e^{-h^2/2}}{\sqrt{2\pi}}\quad 
\label{ph_com_rs}
\end{eqnarray}
and for the {\small PARITY} machine
\begin{eqnarray}
p(h)=\frac{1}{2}\frac{e^{-h^2/2}}{\sqrt{2\pi}} + 
\frac{e^{-h^2/2}}{\sqrt{2\pi}}\:4\!\int\limits_0^{|h|}Dt\:H(t) + 
\frac{1}{12}\delta_{-}(h)+\frac{1}{12}\delta_{+}(h)\quad .
\label{ph_par_rs}
\end{eqnarray}
Note that $p(h)$ for the {\small PARITY} machine is an even function due to the 
additional symmetry of the Boolean function $F$ for this case.

\begin{figure}[htb]
\begin{center}
~
\mbox{\epsfxsize=70mm%
\epsffile[41 50 542 474]{"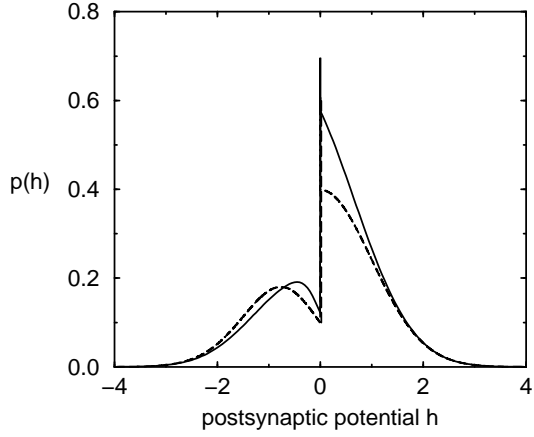"}}
~
\end{center}
\caption[Fig1]{
Distribution of the local field $h$ at the hidden units of a 
$K=3$ {\small COMMITTEE} tree in one-step RSB (bold) and RS (dashed). 
$\delta_{+}(h)$ is represented by adding its weight to the continuous 
part of the curve whereby for a better presentation the RS peak was shifted
slightly to the right.}
\label{fig1}
\end{figure}

\begin{figure}[htb]
\begin{center}
~
\mbox{\epsfxsize=70mm%
\epsffile[41 50 542 474]{"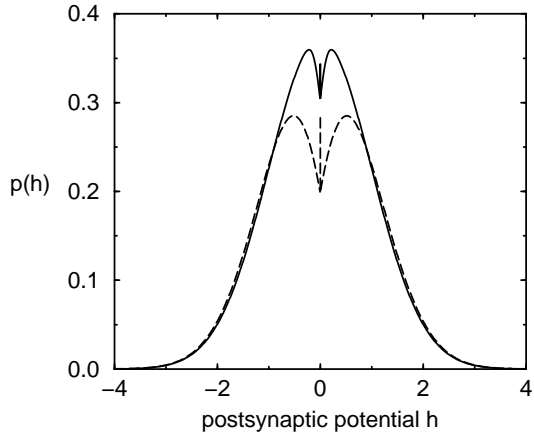"}}
~
\end{center}
\caption[Fig2]{
Distribution of the local field $h$ at the hidden units of a 
$K=3$ {\small PARITY} tree in one-step RSB (bold) and RS (dashed). 
$\delta_{-}(h)$ and $\delta_{+}(h)$ are represented by adding their weights 
to the continuous part of the curve.}
\label{fig2}
\end{figure}

\begin{figure}[htb]
\begin{center}
~
\mbox{\epsfxsize=160mm%
\epsffile[51 89 545 279]{"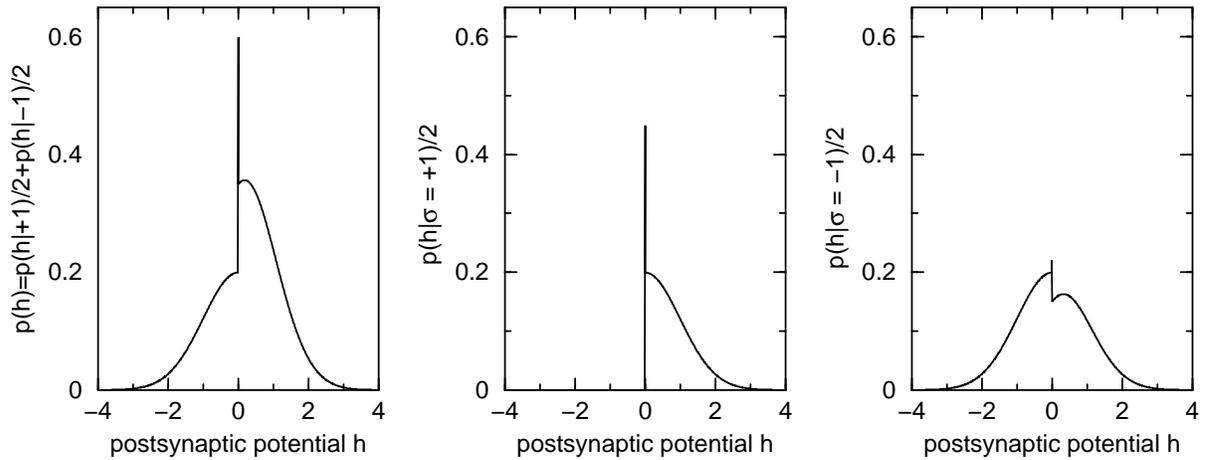"}}
~
\end{center}
\caption[Fig3]{
Distribution $p(h)$ of the local field $h$ at the hidden units of a 
$K=3$ {\small AND} tree in RS (left). The two panels to the right
display its constituents $p(h|\sigma=+1)$ and $p(h|\sigma=-1)$ according
to Eqs. (\ref{ph_andPLUS_rs}) and (\ref{ph_andMIN_rs}).
We found no RSB.
$\delta_{-}(h)$ and $\delta_{+}(h)$ are represented by adding their 
weights
to the continuous part of the curve.}
\label{fig3}
\end{figure}

In the {\small AND} machine the output $\sigma=+1$ can be realized by one LIR only 
whereas the output $\sigma=-1$ results from all the remaining $2^K-1$ IR. 
Hence $p(h|+1)$ and $p(h|-1)$ differ significantly. 
In fact we find for the $K=3$ {\small AND} machine 
\begin{eqnarray}
p(h|+1)&=&\Theta(h)\frac{e^{-h^2/2}}{\sqrt{2\pi}} + \frac{1}{2}\delta_{+}(h),
\label{ph_andPLUS_rs}\\
p(h|-1)&=&\Theta(-h)\frac{e^{-h^2/2}}{\sqrt{2\pi}} 
+\frac{1}{24}\delta_{-}(h) 
+\Theta(h)\frac{e^{-h^2/2}}{\sqrt{2\pi}}(1-H^2(h)),  
\label{ph_andMIN_rs}
\end{eqnarray}
and $p(h)=[p(h|+1)+p(h|-1)]/2$.
Note that we have introduced two different singular contributions 
$\delta_{-}(h)$ and $\delta_{+}(h)$ in 
Eqs. (\ref{ph_com_rs}), (\ref{ph_par_rs}) and 
Eqs. (\ref{ph_andPLUS_rs}), (\ref{ph_andMIN_rs}). 
The reason for this is that the weight of $\delta_{+}(h)$ adds to the 
probability of positive local fields whereas the weight of $\delta_{-}(h)$ 
adds to that of negative local fields. This distinction will be important 
later
when calculating the correlation coefficients from $p(h)$ (cf. Eq. (\ref{h1})).
The results (\ref{ph_com_rs}), (\ref{ph_par_rs}) and (\ref{ph_andPLUS_rs}),
(\ref{ph_andMIN_rs}) are shown as the dashed lines in  
Figs. \ref{fig1}-\ref{fig3} respectively.

These RS results are in fact very intuitive and can be even quantitatively 
understood by assuming that the outcome of a Gardner calculation 
corresponds to the result of a learning process in which the 
initially wrong IR are eliminated with least adjustment \cite{En96}. 
Due to the permutation symmetry between the hidden units we may consider 
only the local field $h_1$ of the first unit of the 
hidden layer. Before learning the couplings $\J_k$ are uncorrelated with the 
patterns and the local field $h_1$ is consequently Gaussian distributed with 
zero mean and unit variance.

Now consider, e.g., the {\small PARITY} machine. Due to the discussed symmetries it 
is sufficient to analyze the case $\sigma=+1$ and $h_1>0$.
If $h_2$ and $h_3$ are equal in sign, which will occur with probability 
$1/2$, there is no need to modify the couplings at all. This gives rise to 
the first term in Eq. (\ref{ph_par_rs}) which is just the original Gaussian and 
describes the chance that a randomly found IR with $h_1>0$ is legal. 
If $h_2$ and $h_3$ differ in sign the IR is illegal and the couplings $\J_k$ 
have to be modified until one of the hidden units changes sign. In an optimal
learning scenario the local field with the smallest magnitude would be 
selected and the corresponding coupling vector would be modified such  
that the field just barely changes sign. Hence $h_1$ remains still unmodified
if either $h_2$ or $h_3$ is smaller in absolute value which gives rise to the
second term in Eq. (\ref{ph_par_rs}). Finally, if really $h_1$ is selected for 
the 
sign change, which will happen with probability $1/6$ for symmetry reasons, it
will after learning be either slightly smaller or slightly larger than zero, 
which is the origin of the last two terms in Eq. (\ref{ph_par_rs}).

With a similar reasoning it is possible to rederive the RS result for the 
{\small COMMITTEE} machine. Again it is sufficient to consider the case $\sigma=+1$.
If $h_1>0$ initially it will not be modified, which gives rise to the last 
term
in Eq. (\ref{ph_com_rs}). If, on the other hand, $h_1<0$, prior to learning it 
will not be modified only if both $h_2$ and $h_3$ are either positive from 
the
start or easier to make positive than $h_1$. Hence a negative $h_1$ survives 
the learning process if 
the other two fields are both larger. This is described by 
the first term in Eq. (\ref{ph_com_rs}). Finally, with probability $5/24$ we find
that $h_1<0$ and either $h_2$ or $h_3$ is even smaller than $h_1$ and 
therefore
harder to correct. In this case the learning would shift $h_1$ to positive 
values as described by the second term in Eq. (\ref{ph_com_rs}). The resulting 
distribution of local fields will hence have a dip for negative values of 
small absolute value clearly visible in Fig. \ref{fig1}.

The case of the {\small AND} machine is the simplest.  
The output $\sigma=+1$ requires all local fields to be positive. 
Hence positive fields are not modified, negative ones are shifted to $0^+$ 
resulting immediately in Eq. (\ref{ph_andPLUS_rs}) which is, of course, 
identical to the result for the single-layer perceptron \cite{Ga89,Op89}.
In the case of a negative output $\sigma=-1$ only the IR $(+,+,+)$ is 
illegal and must be eliminated which is again done by changing the sign of 
the smallest field. This gives rise to Eq. (\ref{ph_andMIN_rs}).

It is finally interesting to compare the distribution of local fields found 
above with that for a single perceptron above saturation \cite{Amit,PMaj}.
The individual perceptrons in a MLN certainly operate above their storage 
limit
even when the storage capacity of the MLN is not yet reached. The most 
remarkable feature of the distribution of local fields for a perceptron above
saturation minimizing the number of misclassified inputs is a {\it gap} 
separating positive from negative values. Being 
intimately related to the failure of any finite level of RSB for this problem
this gap is believed to exist even in the solution with continuous RSB 
\cite{GyRe}. On the other hand, none of the distributions for MLN showed a 
gap.

As should be clear from the above qualitative discussion the reason for this 
is quite simple. The single perceptron above saturation has to reject some 
inputs as not correctly classifiable. In order to keep the number of these 
errors smallest it chooses those with negative fields of large absolute 
value.
Inputs with initially only slightly negative local fields will be learned 
whereby their local fields shift to values just above zero. In this way the 
gap occurs. In MLN, on the other hand, there is no reason to shift all negative
local fields of small absolute value because the correct output may be 
realized by the other hidden units. Therefore one will not find an interval 
of $h$ values for which $p(h)$ is strictly zero. On the other hand, the 
tendency 
that predominantly fields of small absolute value will be modified in the 
learning process is clearly shown by the dips of the distribution functions 
around $h=0$ (cf. Figs. \ref{fig1}-\ref{fig3}).

\subsection{Replica symmetry breaking}
Let us now discuss how the above results get modified by RSB. 
The analytical and subsequent numerical analysis of 
Eqs. (\ref{ph_rsb}-\ref{saddle2}) for 
the $K=3$ machines 
under consideration needs some care in order not to miss the various singular
contributions. We have first to determine the values of the order parameters 
at the saddle point using Eq. (\ref{saddle2}).
In the saturation limit $q_1\to 1$, $(\Phi_{\rm LIR}(\sigma))^m$ is 
dominated by one specific LIR which is selected among all other LIR by the 
sign and absolute value of the compound variables 
$v_k=y_k\sqrt{q_0}+z_k\sqrt{q_1-q_0}$. 
$(\Phi_{\rm LIR}(\sigma))^m$ either tends to 1 or becomes exponentially small in 
one or more compound variables $v_k$. Transforming the integration from 
$z_k$ space to $v_k$ space allows us to reduce the $K$-fold $z$ integral 
to a one-dimensional integral. This is performed numerically by 
Rhomberg integration whereas the outer $y_k$ integrals are done using
Gauss-Legendre quadrature \cite{numrec}.

The saddle point equation (\ref{saddle2}) is solved with a standard 
minimization routine (Powells method in two dimensions \cite{numrec}).
The values we get for the order parameters and for the storage capacity 
are consistent with those obtained earlier. 
For the $K=3$ {\small PARITY} machine we find $q_0=0$, $w\simeq 67.2$, and 
$\alpha^{\rm RSB}_c\simeq 5$ in agreement with\cite{par}. In the case of the
$K=3$ {\small COMMITTEE} machine we get $q_0\simeq 0.64$, $w\simeq 21.2$, and 
$\alpha^{\rm RSB}_c\simeq 3.14$, a result somewhat larger than reported 
previously \cite{Barkai,En92}. The $K=3$ {\small AND} machine finally does not show 
RSB at all and we find accordingly $q_0\rightarrow 1$, $w\rightarrow \infty$ 
together with $\alpha^{\rm  AND}_c=1.31$.

In a second step, we use this values of the order parameters 
$w,\:q_0$ to calculate the respective distribution of local 
fields (\ref{ph_rsb}). 
The distribution functions $p(h)$ obtained in this way are included as full 
lines in Figs. \ref{fig1}-\ref{fig3}. 
Table \ref{tab1} quantifies the main changes.  
The main modification of the distribution functions of local fields 
that occurs in one-step RSB is a redistribution of probability from the 
$\delta$ peaks at $h=\pm 0$ to the continuous part of the distribution around
zero resulting in a reduction of the weight of the singular parts of roughly 
50\%. This gives rise to a less pronounced dip of the distribution functions
around $h=0$ and is qualitatively similar to the RSB modifications for 
a single perceptron above saturation \cite{PMaj}. From the results for the
{\small PARITY} machine it is conceivable that the central peak may get reduced 
further if higher orders of RSB are included and that it might eventually 
disappear completely in the full Parisi solution using continuous RSB.
For all machines the probability of fields with large absolute values is 
hardly affected by RSB.

For the {\small AND} machine we did not find RSB at all. The numerical solution of the
saddle point equations only gave the RSB result $q_0=1$, 
$w\rightarrow\infty$. We therefore suspect that replica symmetry is correct
for the {\small AND} machine. This is also in accordance with the 
rule of thumb that RSB is necessary if the solution space is 
disconnected.
In the {\small AND} machine the output $\sigma=+1$ can be realized
only by one LIR which clearly corresponds to a connected (even convex) 
solution space. The output $\sigma=-1$ is realized by all remaining IR, 
which as the complement of the previous solution space must be connected 
too. 

We have finally to clarify how much the modifications found for the 
distributions of local fields will change the probabilities of the internal 
representations and the correlation coefficients $c_n$ depending only on the 
{\it sign} of the local fields. 

This question is, in fact, nontrivial only in 
the case of the {\small COMMITTEE} machine. For the {\small AND} machine no RSB occurs at all 
and for the {\small PARITY} machine the correlation coefficients are completely 
determined by the symmetry of the Boolean function $F$ between hidden units 
and output.

For the {\small COMMITTEE} machine we find that the probability of the LIR $(+,+,+)$ 
is
shifted from its RS value $0.1250$ to $0.1417$, which is an increase by roughly 
13\% whereas the probability of the three remaining LIR (consisting of two 
pluses and one minus each) is reduced by 1.9\% from 0.2917 to 0.2861.
Qualitatively this means that more inputs are stored with the LIR $(+,+,+)$ 
than the fraction $1/8$ that had this LIR already by chance before learning. 
The learning process hence does not shift illegal IR just up to the decision 
boundary of the Boolean $F$ but in some cases the correlations between inputs
$\ksi$ and couplings $\J$ neglected in RS allow even the safer LIR $(+,+,+)$.

Using Eq. (\ref{corr1}) we can now also calculate the correlation coefficients 
and
find that $c_1$ increases by 2.7\% from its RS value 5/12, $c_2$ decreases 
in absolute value by
13.3\% from its RS value --1/6 and $c_3$ decreases in absolute value 
by 4.5\% from its RS value 
--3/4. This confirms the prediction of \cite{En96} that although crucial 
for the storage capacity RSB will have only a minor influence on the 
correlation coefficients in MLN.

%++++++++++++++++++++++++++++++++++++++++++++++++++++++++++++++++++++++++++++
\section{Summary}
%++++++++++++++++++++++++++++++++++++++++++++++++++++++++++++++++++++++++++++
Generalizing the calculation of the distribution function of local fields for
the single-layer perceptron we introduced a general formalism to determine 
the joint probability distribution $p(h_1,...,h_K)$ of local fields at the 
$K$ hidden units of a two-layer neural network of tree architecture with 
fixed Boolean function between hidden layer and output both in replica 
symmetry
and in one-step replica symmetry breaking. Explicit results were 
obtained for the {\small PARITY}, {\small COMMITTEE}, and {\small AND} machine with $K=3$ hidden units 
in the saturation limit $\alpha\to \alpha_c$. 
Although the individual perceptrons are by far overloaded there is no gap 
in the distribution of local fields as known from a single perceptron above 
saturation. There is no RSB for the {\small AND} machine which we attribute to the 
connected solution space for this architecture. For the {\small PARITY} and 
{\small COMMITTEE} machine we find as a result of RSB a slight redistribution of 
probability from the singular parts at $h=\pm 0$ to the continuous part 
around the origin. The correlation coefficients $c_n$ characterizing the 
correlations between the legal internal representations are not modified by 
RSB for the {\small PARITY} machine since in this case they are fixed already by 
symmetries. For the {\small COMMITTEE} machine the changes of the correlation 
coefficients are rather small and the RS results derived in \cite{En96} may 
serve as useful approximations.

%++++++++++++++++++++++++++++++++++++++++++++++++++++++++++++++++++++++++++++
\appendix
\section*{Replica calculation}
%++++++++++++++++++++++++++++++++++++++++++++++++++++++++++++++++++++++++++++
In this appendix we give some more details on the calculation of the 
distribution function $p(h)$ of the local fields at the hidden units
following Gardner's approach \cite{Ga89}.

Introducing the replica trick 
$1/{\cal Z}=\lim_{n\rightarrow 0}{\cal Z}^{n-1}$ into Eq. (\ref{p_h}) yields
\begin{eqnarray}
p(h)&=& \lim\limits_{n\rightarrow 0}
\bigg\langle\!\!\bigg\langle 
\int \prod\limits_{k,a}d\mu(\J_k^a)d\mu(\lambda^{\nu,a}_k)
\prod\limits_{\nu,a}
\Theta\left(\sigma^{\nu} F(\{\sgn(\lambda^{\nu,a}_k)\})\right)
\delta\left(h-\lambda^{1,a}_1\right)
\bigg\rangle\!\!\bigg\rangle_{\{\ksi_k^{\nu}\},\sigma^{\nu}},
\label{ph_replic1}
\end{eqnarray}
with replica index $a=1,\dots,n$.  
In the integration measures (\ref{measure_J}), (\ref{measure_fields}) we 
replace
the $\delta$ functions by their integral form
\begin{eqnarray}
\delta\left((\J_k^a)^2 - \frac{N}{K}\right)=\int \frac{dE^a_k}{4\pi}
\exp\left(-\frac{i}{2}E^a_k\left((\J_k^a)^2 - \frac{N}{K}\right)\right),
\end{eqnarray}
\begin{eqnarray}
\delta\left(\lambda^{\nu,a}_k-\J^a_k\ksi^{\nu}_k\sqrt{\frac{K}{N}}\right)=
\int \frac{dx_k^{\nu,a}}{2\pi}
\exp\left(ix_k^{\nu,a}\left(\lambda^{\nu,a}_k-\J^a_k\ksi^{\nu}_k
\sqrt{\frac{K}{N}}\right)\right).
\end{eqnarray}

We now perform the average over the Gaussian distributed patterns 
$\xi^{\nu}_{k,i},\:i=1,\dots,N/K$ and introduce the overlaps
$q^{ab}_k=\J^a_k \J^b_k/(N/K)$
of different replicas of the 
same perceptron $\J_k$ as well as its 
conjugated variable $F^{ab}_k$. 
From the assumed permutation symmetry of the Boolean function $F$ with 
respect to all hidden units we infer 
$q^{ab}_k=q^{ab}$, 
$F^{ab}_k=F^{ab}$, and 
$E^a_k=E^a$ for all $k=1,\dots,K$.
This gives rise to the form
\begin{eqnarray}
p(h)=\lim\limits_{n\rightarrow 0}
\int\prod\limits_{a}\frac{dE^a}{4\pi}\!\!
\int\!\!\prod\limits_{a<b} \frac{dq^{ab}dF^{ab}}{2\pi (K/N)}
\:\langle\!\langle p(h|\sigma)\rangle\!\rangle_{\sigma}\:
\exp\left(
\frac{N}{2}{\rm tr}(QA)+
(\alpha N-1)\langle\!\langle \ln G_1(Q|\sigma)\rangle\!\rangle_{\sigma}
+N\ln G_2(A)
\right)
\label{ph_replic2}
\end{eqnarray}
where $Q$ and $A$ denote the symmetric matrices  
$Q^{aa}=1$, $Q^{a\ne b}=q^{ab}$ and $A^{aa}=iE^{a}$, 
$A^{a\ne b}=-iF^{ab}$. Moreover,
\begin{eqnarray}
p(h|\sigma)&=&{\displaystyle \int}\prod_{k,a} 
\frac{d\lambda_k^{a}dx_k^{a}}{2\pi}
\exp\left(\sum_{k,a}\left[i\;x_k^{a}\lambda_k^{a}-\frac{1}{2}
(x_k^{a})^2\right]
-\sum_{k,a<b}x_k^{a}x_k^{b}q^{ab}\right)
\nonumber\\
&\times&\prod\limits_{a}\Theta(\sigma
F(\{\sgn(\lambda_k^{a})\}))\delta(h-\lambda_1^{1}),
\end{eqnarray}
\begin{eqnarray}
G_1(Q|\sigma)&=&{\displaystyle \int}\prod_{k,a} 
\frac{d\lambda_k^{a}dx_k^{a}}{2\pi}
\exp\left(\sum_{k,a}\left[i\;x_k^{a}\lambda_k^{a}-\frac{1}{2}
(x_k^{a})^2\right]
-\sum_{k,a<b}x_k^{a}x_k^{b}q^{ab}\right)
\nonumber\\
&\times&\prod\limits_{a}\Theta(\sigma
F(\{\sgn(\lambda_k^{a})\})),
\label{G1}\\
G_2(A)&=&\int\!\prod\limits_{k,a}\frac{dJ_k^a}{\sqrt{2\pi e}}
\exp\left(-\frac{1}{2}\sum\limits_{k;a,b} J_k^aA^{ab}J_k^b\right).
\end{eqnarray}

In the limit $N\rightarrow \infty$ the integral in Eq. (\ref{ph_replic2}) is 
dominated by the saddle point values of the order parameters $E_a$, $F^{ab}$, 
and $q^{ab}$.
Solving the saddle point equation with respect to $E^{a}$ and $F^{ab}$ 
yields $A=Q^{-1}$.
Hence Eq. (\ref{ph_replic2}) takes the form
\begin{eqnarray}
p(h)=\lim\limits_{n\rightarrow 0}
\int\prod\limits_{a<b} 
dq^{ab}\:\langle\!\langle p(h|\sigma)\rangle\!\rangle_{\sigma}
\:
\exp\left(\frac{N}{2}
\ln {\rm det}(Q)
+(\alpha N-1)\langle\!\langle\ln G_1(Q|\sigma)\rangle\!\rangle_{\sigma}
\right)
.
\label{ph_replic3}
\end{eqnarray}
$p(h|\sigma)$ can be calculated by assuming either RS or one-step RSB for 
the matrix $Q$ resulting in Eqs. (\ref{ph_rs}) and (\ref{ph_rsb}), 
respectively.

The remaining saddle point condition for the matrix $q^{ab}$ 
has in one-step RSB (\ref{rsb}) the form
\begin{eqnarray}
&{\rm extr}_{q_0,q_1,m}&\left[
\frac{q_0}{2[1-q_1+m(q_1-q_0)]} + 
\frac{1}{2m}\ln\left(1+\frac{m(q_1-q_0)}{1-q_1}\right)
+\frac{1}{2}\ln(1-q_1)\right.
\nonumber\\
&+&\left.\frac{\alpha}{m}
\left\langle\!\!\left\langle\int\prod\limits_{k}Dy_k\ln\left\{
\int\prod\limits_{k}Dz_k\left(\Phi_{\rm LIR}(\sigma)\right)^m\right\}
\right\rangle\!\!\right\rangle_{\sigma}
\right].
\label{saddle1}
\end{eqnarray}

It determines a set of order parameters $\{q_1,q_0,m\}$ for every pattern 
load below the storage capacity $\alpha\le\alpha_c$. 
The abbreviation $\Phi_{\rm LIR}(\sigma)$ is defined by Eq. (\ref{sum_LIR_1}).
The angular brackets $\langle\langle\cdots\rangle\rangle_{\sigma}$ indicate 
the average over the two possible outputs $\sigma=\pm1$.

%++++++++++++++++++++++++++++++++++++++++++++++++++++++++++++++++++++++++++++
%++++++++++++++++++++++++++++++++++++++++++++++++++++++++++++++++++++++++++++
%++++++++++++++++++++++++++++++++++++++++++++++++++++++++++++++++++++++++++++

%++++++++++++++++++++++++++++++++++++++++++++++++++++++++++++++++++++++++++++

\begin{table}[htb]
\caption[Tab1]{
Saturated $K=3$ machines: Integrated features of the probability 
distribution $p(h)$ of the local field. 
Corrections by one-step RSB are given in percent of the respective RS value. 
Dashes indicate that a respective singular contribution does not occur
({\small COMMITTEE}) or that we found no RSB ({\small AND}).}
\label{tab1}
\begin{center}
~
\begin{tabular}{l|c|c|c|c}
& Negative non-
& Singular
& Singular
& Positive non-\\
& zero fields
& contribution
& contribution
& zero fields\\
& $\lim_{\epsilon \rightarrow 0}\int_{-\infty}^{-|\epsilon|}
\! p(h)\:dh$
& $\delta_{-}(h)$
& $\delta_{+}(h)$
&  $\lim_{\epsilon \rightarrow 0}\int_{|\epsilon|}^{\infty}
\! p(h)\:dh$\\ 
\begin{tabular}{l} 
$K=3$ machine \hspace*{0.4cm}\\       \hline
{\small COMMITTEE}    \\       \hline
{\small PARITY}       \\       \hline 
{\small AND}
\end{tabular}
&
\begin{tabular}{ll}
\hspace*{0.4cm}RS $\:\:\:\:\:\:\:\:\:\:\:\:$      & 1-RSB    
\\ \hline
\hspace*{0.4cm}$7/24$       & $-$1.9$\%$  \\ \hline
\hspace*{0.4cm}$5/12$       & $+$10.3$\%$\hspace*{0.4cm} \\ \hline
\hspace*{0.4cm}$1/4$        & $-$       
\end{tabular}
&
\begin{tabular}{ll}
\hspace*{0.4cm}RS $\:\:\:\:\:\:\:\:\:\:\:\:$       & 1-RSB     
\\ \hline 
\hspace*{0.4cm}$-$          & $-$        \\ \hline
\hspace*{0.4cm}$1/12$       & $-$51.8$\% $\hspace*{0.4cm} \\ \hline
\hspace*{0.4cm}$1/48$       & $-$        
\end{tabular}
&
\begin{tabular}{ll}
\hspace*{0.4cm}RS $\:\:\:\:\:\:\:\:\:\:\:\:$       & 1-RSB     
\\ \hline 
\hspace*{0.4cm}$5/24$       & $-$42.3$\% $ \\ \hline
\hspace*{0.4cm}$1/12$       & $-$51.8$\% $\hspace*{0.4cm} \\ \hline
\hspace*{0.4cm}$1/4$        & $-$        
\end{tabular}
&
\begin{tabular}{ll}
\hspace*{0.5cm}RS $\:\:\:\:\:\:\:\:\:\:\:\:$      & 1-RSB   
\\ \hline 
\hspace*{0.5cm}$1/2$        & $+$18.9$\%$ \\ \hline
\hspace*{0.5cm}$5/12$       & $+$10.3$\%$\hspace*{0.4cm} \\ \hline
\hspace*{0.5cm}$23/48$      & $-$   
\end{tabular}
\end{tabular}
~
\end{center}
\end{table}

%++++++++++++++++++++++++++++++++++++++++++++++++++++++++++++++++++++++++++++
%++++++++++++++++++++++++++++++++++++++++++++++++++++++++++++++++++++++++++++
\newpage
%__________________
%___________________ 

%__________________
%___________________

%__________________
%___________________

\vspace{2cm}
%++++++++++++++++++++++++++++++++++++++++++++++++++++++++++++++++++++++++++++
%++++++++++++++++++++++++++++++++++++++++++++++++++++++++++++++++++++++++++++
%++++++++++++++++++++++++++++++++++++++++++++++++++++++++++++++++++++++++++++

\end{document}